\documentclass[twocolumn,superscriptaddress,prl]{revtex4-1}
\usepackage{graphicx}
\usepackage{amsmath}
\usepackage{amsthm}
\usepackage{amssymb}
\usepackage{latexsym}
\usepackage{array}
\usepackage{float}
\usepackage{amsfonts}
\usepackage{mathrsfs}
\usepackage{verbatim}
\usepackage{color}
\usepackage[capitalize]{cleveref}

\usepackage{soul}

\usepackage[bb=boondox]{mathalfa}

\usepackage{makecell}
\usepackage{adjustbox,lipsum}

\begin{document}

\title{Experimental Learning of Pure Quantum States using Sequential Single-Shot Measurement Outcomes}

\author{Sang Min Lee}\email{samini@kriss.re.kr}
\affiliation{Korea Research Institute of Standards and Science, Daejeon 34113, Korea}

\author{Hee Su Park}
\affiliation{Korea Research Institute of Standards and Science, Daejeon 34113, Korea}

\author{Jinhyoung Lee}
\affiliation{Department of Physics, Hanyang University, Seoul 04763, Korea}

\author{Jaewan Kim}
\affiliation{School of Computational Sciences, Korea Institute for Advanced Study, Seoul 02455, Korea}

\author{Jeongho Bang}\email{jbang@kias.re.kr}
\affiliation{School of Computational Sciences, Korea Institute for Advanced Study, Seoul 02455, Korea}

\received{\today}
\newcommand{\bra}[1]{\left<#1\right|}
\newcommand{\ket}[1]{\left|#1\right>}
\newcommand{\abs}[1]{\left|#1\right|}
\newcommand{\expt}[1]{\left<#1\right>}
\newcommand{\braket}[2]{\left<{#1}|{#2}\right>}
\newcommand{\commt}[2]{\left[{#1},{#2}\right]}
\newcommand{\tr}[1]{\mbox{Tr}{#1}}

\newcommand{\blue}[1]{\textcolor{blue}{#1}}
\newcommand{\red}[1]{\textcolor{red}{#1}}
\newcommand{\green}[1]{\textcolor{green}{#1}}

\newcommand{\itbf}[1]{\textit{\textbf{#1}}}
\newcommand{\rem}[1]{\red{\st{#1}}}

\begin{abstract}
We experimentally implement a machine-learning method for accurately identifying unknown pure quantum states. The method, called single-shot measurement learning, achieves the theoretical optimal accuracy for $\epsilon = O(N^{-1})$ in state learning and reproduction, where $\epsilon$ and $N$ denote the infidelity and number of state copies, without employing computationally demanding tomographic methods. This merit results from the inclusion of weighted randomness in the learning rule governing the exploration of diverse learning routes. We experimentally verify the advantages of our scheme by using a linear-optics setup to prepare and measure single-photon polarization qubits. The experimental results show highly accurate state learning and reproduction exhibiting infidelity of $O(N^{-0.983})$ down to $10^{-5}$, without estimation of the experimental parameters.
\end{abstract}

\maketitle


{\em Introduction.}---Recently, there has been increasing interest in applying machine learning to quantum information tasks~\cite{Carleo2019}. This is apparent from the increased use of, for example, state identification and tomography methods based on a Bayesian model~\cite{Huszar2012}, neural networks~\cite{Torlai2018,Palmieri2020}, and other learning approaches~\cite{Mahler2013,Ferrie2014,Qi2017}. These machine-learning-based methods can retrieve critical information on the quantum state through stepwise data processing, even without \textit{a priori} knowledge of the state. Importantly, the optimal accuracy can be achieved with Heisenberg-limited scaling, namely, the infidelity of $O(N^{-1})$ for a finite number $N$ of resources~\cite{Kravtsov2013,Lee2018}.

Experimental implementation of learning-based state identification methods is generally challenging since the achievable accuracy is limited by imperfections in the learning apparatus, say ${\cal L}$. A set of measurement data ${\cal D} = \{ m(\{\alpha\}_{\cal L}) \}$ for experimental parameters $\{\alpha\}_{\cal L}$ are mapped to new parameters $\{\alpha^{\text{new}}\}_{\cal L}$ in each learning step. Algorithms determine the procedure to choose the next $\alpha^{\text{new}}$ during iterations. Under realistic conditions, the measurement datasets are affected by systematic errors~\footnote{For example, systematic errors can arise due to non-ideal $\pi/2$ or $\pi$-pulse in (quasi-)atomic system. In optical system, non-exact phase retardation of waveplates and finite extinction ratio of polarizers can cause these errors.} and the resulting decision system shows the effect of accumulated errors. When original unknown states need to be reproduced, for example, in some cryptographic tasks~\cite{Liang2003,Bogdanov2010}, their accuracy is further reduced by imperfections in the reproduction setup components.

We experimentally implement an error-robust and generally applicable learning algorithm, called single-shot measurement learning (SSML), in which decision-making is based on sequential single-shot measurement outcomes~\cite{Lee2018}. The algorithm involves trial operator variations with weighted amounts of randomness. Conceptually, the magnitude of the random variation decreases according to the number of consecutive {\em success} events before encountering a {\em failure} event, after which the variation of the learning operator is applied to the experimental setup. The SSML estimator is applicable to arbitrary unknown pure states with minimal free parameters, and it can achieve $O(N^{-1})$ accuracy without procedures requiring extensive computational loads. Our algorithm adaptively identifies the unitary operator to inter-convert between a fixed initial state and an unknown state, and can reproduce highly accurate copies of the unknown state despite imperfections in the practical components. The experimental demonstration uses polarization qubits of single photons, $\ket{0}=\ket{H}$ (horizontal) and $\ket{1}=\ket{V}$ (vertical). The measured infidelities scale as $O(N^{-0.983})$ on average down to the accuracy level of $<10^{-5}$.

{\em Method.}---Our algorithm learns a unitary $\hat{U}$ that maps an unknown state $\ket{\psi}$ to a known fiducial state $\ket{\mathbb{0}}$; $\ket{\psi}$ is identified as $\ket{\psi} \simeq \ket{\psi_\text{est}} = \hat{U}^\dagger\ket{\mathbb{0}}$. The basic building blocks of the algorithm are preparation ($\mathbf{P}$), operation ($\mathbf{U}$), measurement ($\mathbf{M}$), and feedback ($\mathbf{F}$). The fiducial state $\ket{\mathbb{0}}$ is freely chosen as the most accurately detectable state. $\mathbf{U}$ and $\mathbf{M}$ constitute the learning apparatus ${\cal L}$, and $\mathbf{P}$ is regarded as a black box that repeatedly generates $\ket{\psi}$~\footnote{The only given information about $\mathbf{P}$ is the Hilbert-space dimension $d$ of $\ket{\psi}$. Such an assumption is commonly used by estimation problems.}. $\mathbf{U}$ implements an arbitrary unitary operation $\hat{U}(\{ \alpha \}_\mathbf{U})$, where the experimental parameters $\{ \alpha \}_\mathbf{U}$ are updated in each learning step by using the measured data ${\cal D}=\{ m(\{\alpha\}_{\cal L} )\}$. ${\cal D}$ is the output of $\mathbf{M}$, which comprises ``yes-or-no'' questions on the desired target's detection. If $\mathbf{M}$ projects the state onto $\ket{\mathbb{0}}$ in a trial, ``success ($s$)'' is tagged to the outcome $m$; otherwise, $m$ is labeled ``failure ($f$).'' The number of consecutive successes $M_S$ directly indicates the current learning status~\footnote{Let us consider the probability $p$ to have $M_S$ consecutive $s$ signals for a learned state $\ket{\psi_\text{est}}$. With the infidelity being $\epsilon = 1 - \abs{\braket{\psi}{\psi_\text{est}}}^2$, the probability $p$ is $\epsilon(1-\epsilon)^{M_S}$, and the expectation value of the number of successes $\overline{M_S}$ is $\sum_{m = 0}^{\infty } m \epsilon (1-\epsilon)^m = \epsilon^{-1}-1$. We thus have $\epsilon = (1+\overline{M_S})^{-1}$, and can estimate $\epsilon$ by counting $M_S$.}.

\begin{figure*}[t]
\centering
\includegraphics[width=0.85\textwidth]{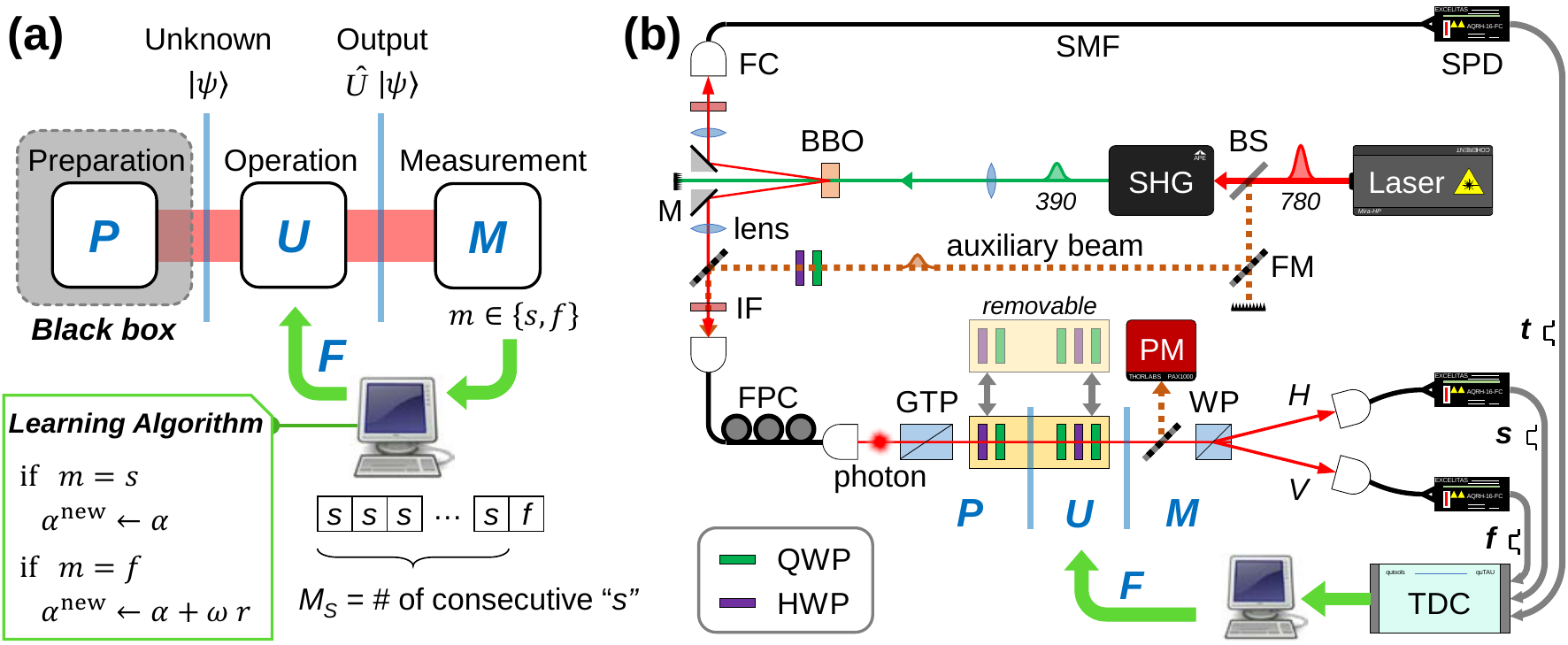}
\caption{(a) Schematic of the SSML for unknown state identification. (b) Experimental setup involving linear optics. $\mathbf{P}$, $\mathbf{U}$, $\mathbf{M}$, and $\mathbf{F}$ denote state preparation, unitary operation, measurement, and feedback, respectively. The auxiliary beam is the fundamental laser light whose center wavelength (780 nm) is the same as the photons. BS: beam-splitter, SHG: second-harmonic generator, BBO: $\beta$-BaB$_2$O$_4$ crystal (thickness $1$ mm), M: mirror, FM: flip mirror, IF: interference filter, FC: fiber coupler, SMF: single-mode fiber, FPC: fiber paddle polarization controller, GTP: Glan-Thompson polarizer, Q(H)WP: rotatable quarter(half)-wave plate, PM: polarimeter, WP: Wollaston prism, $H(V)$: horizontal (vertical) polarization, SPD: single-photon detector, $t$: trigger, $s$: success, $f$ failure, TDC: time-to-digital converter.}
\label{fig:scheme_n_exp}
\end{figure*}

The learning rule for updating $\{ \alpha \}_\mathbf{U}$, which is the set of all experimental parameters used to control $\mathbf{U}$, are as follows: [{\bf F.1}] if $m=s$, we retain $\{ \alpha \}_\mathbf{U}$ and set $M_S \leftarrow M_S +1$, and [{\bf F.2}] if $m=f$, we change each $\alpha$ to $\alpha^{\text{new}} \leftarrow \alpha + \omega r$, where $r$ is a {\em random} number and $\omega = a ( M_S+1 )^{- b}$ is the weight for the random walk; $a$ and $b$ are free parameters chosen to optimize learning performance~\footnote{In the current experiments, we set $a=0.3$, $b=0.5$, and $r \in [-\frac{\pi}{2}, \frac{\pi}{2}]$. The $a$ value was found by trial and error. The $b$ value was chosen by noting that $\epsilon$ is a quadratic function of $\{\alpha\}$ near the optimal solution $\{\alpha_\text{sol}\}$).}. The learning is complete when the halting condition $M_S = M_H$ is reached. From the learned parameters $\{ \alpha_\text{learn} \}_\mathbf{U}$, we identify $\ket{\psi}$ such that $\ket{\psi_\text{est}} = \hat{U}(\{ \alpha_\text{learn} \}_\mathbf{U})^\dagger \ket{\mathbb{0}}$ with a sufficiently small infidelity $\epsilon = 1- \abs{\braket{\psi_\text{est}}{\psi}}^2 \ll 1$. This method is schematically summarized in Fig.~\ref{fig:scheme_n_exp}(a). As the preset $M_H$ increases, the number of state copies $N$ increases and the final $\epsilon$ decreases. The resulting trade-off relation between $N$ and $\epsilon$ determines the overall learning efficiency. 

{\em Experiments.}---Figure~\ref{fig:scheme_n_exp}(b) shows the experimental setup. We prepare heralded single photons through type-I spontaneous parametric down-conversion (SPDC) in a BBO ($\beta$-BaB$_2$O$_4$) crystal pumped by mode-locked laser pulses (wavelength $390$ nm, repetition $76$ MHz, pulse width $150$ fs, average power $35$ mW). The down-converted photons are filtered by interference filters (half-maximum bandwidth $3$ nm) and coupled into single-mode fibers. The upward-propagating photons are detected by a trigger single-photon detector (SPD), and the downward-propagating photons are initialized by $\mathbf{P}$ and transformed by $\mathbf{U}$ before being measured by $\mathbf{M}$. The data are recorded as coincidence counts (time window $\simeq 2$ ns) of the two photons to minimize the effects of detector dark counts and stray light. An arbitrary polarization qubit $\ket{\psi}$ is prepared (part $\mathbf{P}$) using a Glan-Thompson polarizer (extinction ratio $>10^5$), a half-wave plate (HWP), and a quarter-wave plate (QWP). Another QWP-HWP-QWP set constitutes a unitary operator (part $\mathbf{U}$), and the rotation angles $\{ \alpha_1, \alpha_2, \alpha_3 \}_\mathbf{U}$ of the three wave plates are the learning parameters updated according to [{\bf F.1}] and [{\bf F.2}]. The photons are finally split by a Wollaston prism (extinction ratio $>10^5$) into horizontal- and vertical-polarization modes, which are detected by two low-noise SPDs (Excelitas SPCM-AQRH-16). We define the detection of horizontal (vertical) polarization $\ket{H}$ ($\ket{V}$) as success $s$ (failure $f$) (part $\mathbf{M}$). Procedures $\mathbf{U}$ and $\mathbf{M}$ are repeated with $\omega$-weighted random feedback to [{\bf F.2}] until the halting condition is satisfied.

\begin{figure}[t]
\centering
\includegraphics[width=0.48\textwidth]{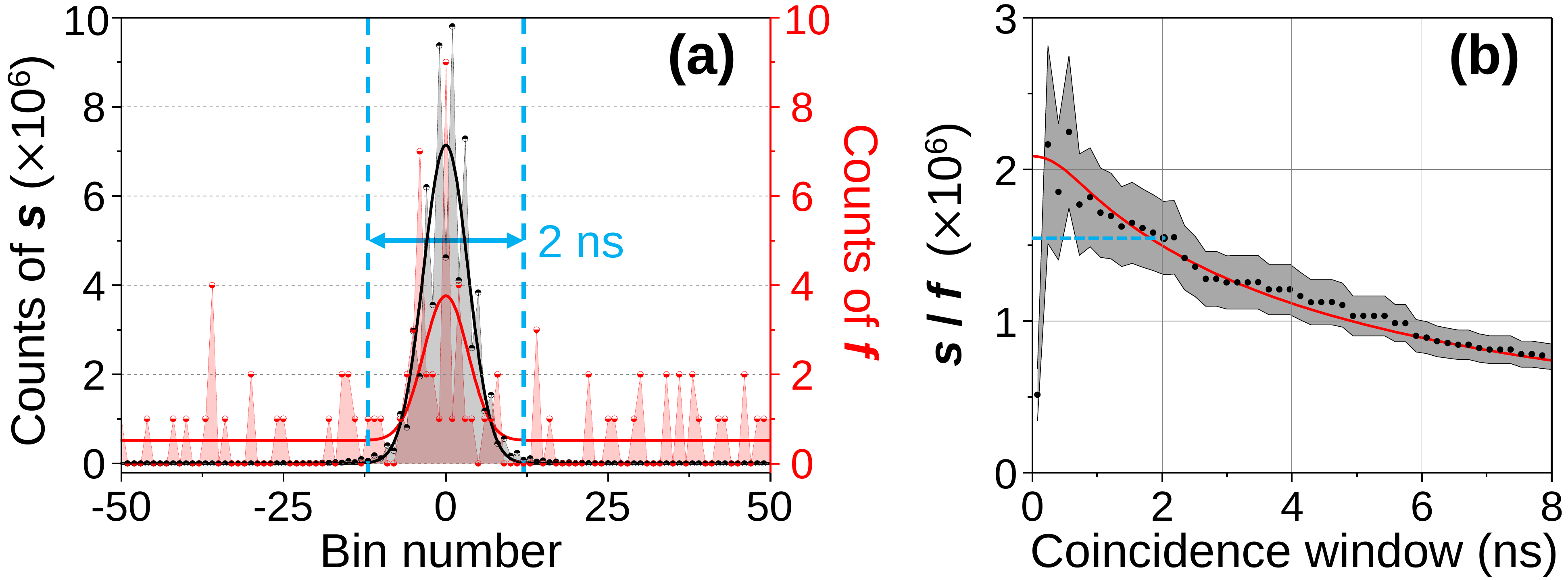}
\caption{\label{fig:SNR} (a) Arrival time distribution of output photons relative to the trigger signal (bin size is about $81$ ps, collection time: 10 h, quTAU). Photons are initialized to horizontal polarization and the wave plates in $\mathbf{P}$ and $\mathbf{U}$ are omitted during this measurement. The curves are Gaussian fits to the measured data. The center peak of the $f$ counts shows the (nonideal) extinction ratio of polarizers. (b) Dependence of the SNR of the measurement setup on the coincidence window. The window size (25 bins $\simeq$ 2 ns) sets an SNR of $1.55(24) \times 10^6$.}
\end{figure}

Experimental imperfections in all the procedures of $\mathbf{P}$, $\mathbf{U}$, $\mathbf{M}$ are suppressed to ensure high accuracy of the experiments. Since our scheme depends on every single detection of photons, false negative signals due to detector dark counts or finite extinction ratios of polarizers critically limit the eventual learning accuracy. $M_S$ exceeding the signal-to-noise ratio (SNR) $\sim 1/q$, where $q$ is the probability to have a false negative (failure) signal, cannot be measured because the noises readily interfere the learning step before $M_S$ successes in a row can be collected. To raise the SNR, we first use polarizers with extinction ratio $> 10^5$ and SPDs with a low dark count rate $\simeq11$ cps. The time window size for coincidence detection of a signal photon and a trigger photon is chosen to compromise between the SNR and the overall detection efficiency. Figure~\ref{fig:SNR}(a) shows the temporal distributions of success ($s$) and failure ($f$) signals with respect to trigger signals. For this measurement all the wave plates for $\mathbf{P}$ and $\mathbf{U}$ are removed not to perturb the initial state $\ket{\mathbb{0}}$ ideally leading to the $s$ detector. The ratio between the $s$ and $f$ counts according to the window size is plotted in Fig.~\ref{fig:SNR}(b). As the SNR decreases in the large window size regime, we set the width as 2 ns that is close to the minimum to enclose the main peak in Fig.~\ref{fig:SNR}(a). This yields an SNR ($s/f$) of $1.54(24) \times 10^{6}$.

To evaluate the infidelity $\epsilon = 1- \abs{\braket{\psi_\text{est}}{\psi}}^2$ experimentally, we compare Stokes vectors of the auxiliary classical lights (sub-mW) whose center wavelength is same as the photons. We set $\ket{\psi}=\hat{V}\ket{\mathbb{0}}$, where $\ket{\mathbb{0}}=\ket{H}$ and $\hat{V}$ is given by the pre-aligned HWP and QWP of $\mathbf{P}$. The overlap $\braket{\psi_\text{est}}{\psi}$ after learning runs is $\bra{\mathbb{0}}\hat{U}\hat{V}\ket{\mathbb{0}}$. A polarimeter (Thorlabs PAX1000IR1) measures the Stokes vectors $\vec{S}_{H'}$ and $\vec{S}_H$ of the classical light with and without passing through the wave plates in $\mathbf{U}\mathbf{P}$, respectively. We then obtain $\epsilon = \sin^2{\frac{\phi}{2}}$, where $\phi$ is the angle between $\vec{S}_{H'}$ and $\vec{S}_H$. Notably, $\epsilon$ is evaluated without estimating $\alpha_\text{learn}$ or identifying $\ket{\psi_\text{est}}$. The standard deviation of the direction of Stokes vectors was $2$ mrad in our operation mode (50 Hz, 2048 pts FFT), and each measurement was repeated 100 times. Therefore the accuracy limit of $\epsilon$ owing to our experimental setup is estimated to be about $10^{-8}$~\footnote{The $\epsilon$-estimation limit from uncertainty of polarimeter is $\sin^2\frac{u(\phi)}{2} \simeq \frac{s(\phi)^2}{4n} \simeq \frac{(2 \times10^{-3})^2}{4 \times 100}=10^{-8}$, where $u(\phi)$ and $s(\phi)$ denote the standard uncertainty and the standard deviation, respectively. It can be further improved by increasing the number $n$ of samples}. Note that this direct comparison method avoid the errors caused by non-ideal retardation or angle offsets of the wave plates, in contrast to the evaluation of the infidelity using the experimentally obtained $\alpha_\text{learn}$.

\begin{figure*}[t]
\centering
\includegraphics[width=0.98\textwidth]{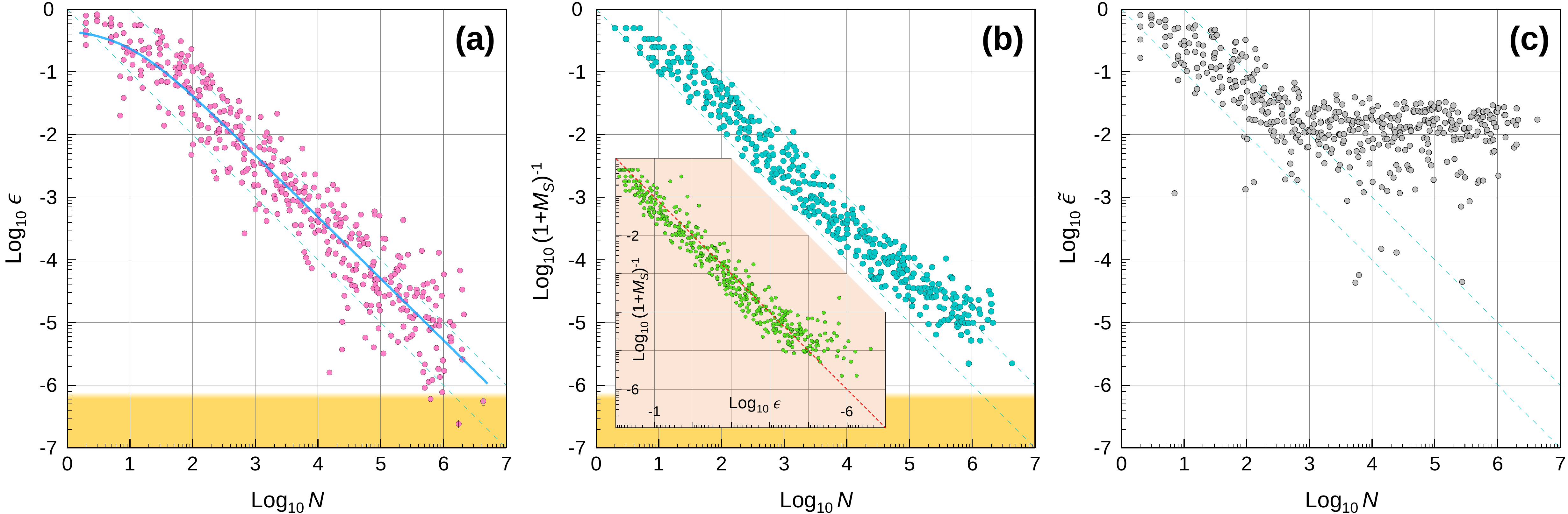}
\caption{\label{fig:EXP_results} Accuracy of state estimation versus the amount of used resources. (a) Independently measured infidelity $\epsilon$. (b) Monitored infidelity $(1+M_S)^{-1}$. (c) Infidelity estimated by the rotation angles of the wave plates. The learning accuracy defined by the number of consecutive success detections matches reasonably well with the measured infidelity as shown in the inset of (b). The (yellow) shaded regions in (a) and (b) denote the measurement limit given by the finite SNR of the current setup. See the main texts for more details.}
\end{figure*}

The experiments were repeated for $35$ random unknown input states. We set $M_H=6 \times 10^4$ as a halting condition. Whenever $M_S$ after detection of an $f$-signal exceeds the previous maximum of $M_S$, $\epsilon$ is experimentally measured and $N$ is recorded as the total copies consumed until that instant. At the moment $M_S$ passes $M_H$, a decision is made to stop the learning at the current $\mathbf{U}$ when the next $f$ is fired. The maximum $M_S$ observed in our experiment in this way was $451216$. The final halting condition was reached after $6$ to $18$ renewals of $M_S$. Data for $N$ and $\epsilon$ are plotted in Fig.~\ref{fig:EXP_results}(a) and they fit a curve $\epsilon = C (N+N_0)^{-\gamma}$ on a log-log scale, with $\gamma \simeq 0.983(19)$ (blue solid line), implying that our learning accuracy was $O(N^{-0.983})$ for $N$ unknown state copies. Remarkably, this tendency is maintained under the level of $10^{-5}$. The standard uncertainty (SU) of $\epsilon$ is smaller than the point size (except for the two at the bottom right). The minimum of observed $\epsilon$ was $2.4(4) \times 10^{-7}$. Our experimental precision excels the previous methods as summarized in Table~\ref{tab:comparison}.

\begin{table}[t]
\centering
\tabcolsep=0.1in
\begin{adjustbox}{max width=\textwidth}
\begin{tabular}{c c c c}
\hline\hline
Method 	& $\gamma$ 	& $\epsilon_\text{min}$ & Dim \\
\hline
ABQT~\cite{Kravtsov2013} 	& 0.98(1) 		& $> 6$$\times$$10^{-5}$		& 2 \\
SAQST~\cite{Mahler2013} 	& 0.90(4) 		& $> 5$$\times$$10^{-5}$		& 2 \\
SGQT~\cite{Chapman2016} 	& NA$^{1)}$ 		& $ 7(2)$$\times$$10^{-3}$	& 2 \\
RAQST~\cite{Qi2017}	 	& NA$^{2)}$ 		& $> 3$$\times$$10^{-4}$		& 2$\times$2 \\
ABQT~\cite{Struchalin2018}  	& 0.703(16) 	& $> 2$$\times$$10^{-3}$		& 9$\times$9 \\
SSML (This work) 			& 0.983(19) 	& $< 1 \times 10^{-5}$ 	& 2 \\
\hline\hline
\end{tabular}
\end{adjustbox}
\caption{\label{tab:comparison} Comparison with the previous learning-based schemes. ABQT: adaptive Bayesian quantum tomography, SAQST: single adaptive quantum state tomography, SGQT: self-guided quantum tomography, RAQST: recursively adaptive quantum state tomography. $\gamma$ is the scaling factor of the accuracy defined as $\epsilon = O(N^{-\gamma})$. $\epsilon_\text{min}$ is the minimum infidelity achieved in the experiment. Dim denotes the Hilbert-space dimension of the unknown state. $^{1)}$Not calculated. $^{2)}$Multiple slopes.}
\end{table}

We next consider state reproduction based on our scheme. Usually, an unknown state is reproduced by sequentially split procedures for identification and reconstruction. However, our SSML method does not require information on the learned state to configure a state preparation unit since $\mathbf{U}$ learns the unitary corresponding to an experimental setup used for transforming a fiducial state $\ket{\mathbb{0}}$ to the unknown state $\ket{\psi}$. Therefore, reproduction is realized directly by applying $\mathbf{U}^{-1}$ to $\ket{\mathbb{0}}$, i.e. sending a photon in $\ket{\mathbb{0}}$ through $\mathbf{U}$ {\em backward}, after the learning. The accuracy of reproduction can be estimated from the $M_S$ value. Figure~\ref{fig:EXP_results}(b) plots the reproduction accuracy $(1+M_S)^{-1}$ on a log-log scale. Figures ~\ref{fig:EXP_results}(a) and ~\ref{fig:EXP_results}(b) agree with each other as shown in the inset. Adoption of $(1+M_S)^{-1}$ as the learning accuracy is useful because the infidelity of $\ket{\psi_\text{est}}$ needs not be independently estimated. This benefits extension to higher dimensions or multiple particles where standard quantum state tomography becomes more costly.

We compare the infidelities simply deduced from the wave plate angles comprising $\mathbf{P}$ and $\mathbf{U}$, as shown in Fig.~\ref{fig:EXP_results}(c). Experimentally, the rotation angles of the wave plates were calibrated using an auxiliary laser light with an accuracy of $0.02^\circ$. The data in Fig.~\ref{fig:EXP_results}(c) show that it is difficult to maintain $\epsilon = O(N^{-1})$ below $10^{-2}$. Considering the rotational accuracy of the wave plates, we expect that a finite accuracy ($<\lambda/300$) and incidence angle sensitivity of phase retardations in the zero-order Quartz wave plates limit the estimation accuracy of the polarization states in this indirect method.

{\em Summary and discussions.}---We have experimentally realized single-shot-based learning of unknown pure states with an unprecedented level of precision. The linear-optics setup has exploited the merits of the proposed scheme over its fullest potential, to our belief. The fidelities of the learned states with various input states were verified by a devised method using auxiliary classical light. Experimentally achieved infidelities decreased as $O(N^{-0.983})$ almost reaching the theoretical accuracy limit, below $10^{-5}$. We have also verified that the learning accuracy calculated by the number of consecutive successes agrees with the independently measured fidelity between the input and output states. These results suggest that adaptive or machine-learning methods operated by shot-by-shot feedbacks can have practical merits for quantum measurement applications at the highest precision regime.


{\em Acknowledgments.}---S.M.L. and H.S.P. acknowledge the support of the R\&D Convergence program of NST of Republic of Korea (CAP-15-08-KRISS), the KRISS project (GP2020-0010-02, -0013-30) and National Research Foundation of Korea (NRF) grants (No. 2019M3E4A1079894). J.K. was supported in part by KIAS Advanced Research Program (CG014604). J.B. was supported by a KIAS Individual Grant (CG061003). J.L. and J.B. acknowledge  the support of NRF grants (2019R1A2C2005504 and NRF-2019M3E4A1079666). J.B. also acknowledge the research project on developing quantum machine learning and quantum algorithm (No.~2019-100) by the ETRI affiliated research institute.


%

\end{document}